\documentclass[singlecolumn,showpacs,preprintnumbers,amsmath,amssymb]{revtex4}

\usepackage{graphicx}
\usepackage{dcolumn}
\usepackage{bm}
\usepackage{subfigure}
\usepackage{epsfig,amsmath}
\usepackage{subfigure}
\usepackage{stmaryrd}
\usepackage{mathrsfs}
\usepackage{pifont}
\usepackage{amsthm}
\usepackage{amssymb}
\usepackage{latexsym}
\usepackage{hyperref}
\usepackage{color}

\begin{document}

\title{An electro-optic waveform interconnect based on quantum interference}

\author{Li-Guo Qin$^1$$\footnote{[Email address]: qinlg@sari.ac.cn}$, Zhong-Yang Wang$^{1}$\footnote{[Email address]: wangzy@sari.ac.cn}, $\&$ Shang-Qing Gong$^{2}$} 
\affiliation {$^1$Shanghai Advanced Research Institute, Chinese Academy of Sciences, Shanghai, 201210, China \nonumber\\
$^2$Department of Physics, East China University of Science and Technology, Shanghai 200237, China}

\date{\today}

\begin{abstract}

The ability to modulate an optical field via an electric field is regarded as a key function of electro-optic interconnects, which are used in optical communications and information processing systems. One of the main required devices for such interconnects is the electro-optic modulator (EOM). Current EOM based on the electro-optic effect and the electro-absorption effect often is bulky and power inefficient due to the weak electro-optic properties of its constituent materials. Here we propose a new mechanism to produce an arbitrary-waveform EOM based on the quantum interference, in which both the real and imaginary parts of the susceptibility are engineered coherently with the superhigh efficiency. Based on this EOM, a waveform interconnect from the voltage to the modulated optical absorption is realised. We expect that such a new type of electro-optic interconnect will have a broad range of applications including the optical communications and network.

\pacs{85.85.+j, 42.50.Gy, 84.30.Qi}

\end{abstract}

\maketitle

The trend toward on-chip and off-chip photonics has lasted for decades \cite{bce,mdar,sab}. Current integrated Silicon photonics technology can support the design of electronics and photonics on the same chip and enable cost-efficient optical links that connect racks, modules and chips together \cite{sab,sun,vya}. This indicates the beginning of an era of chip-scale electronic-photonic systems \cite{sun}. However, to realise the electro-optic interconnect, the electro-optic chip faces the challenges including the electro-optic transformation and low power consumption \cite{bce}, etc. Moreover, the synchronous interconnect of the electro-optic signals meets the difficulties of the match in the propagation velocity, the high modulation capability and the response speed. To realise the dynamical transformation from the electrical signals to optical ones, the electro-optic modulator (EOM) is a potential device as a main required function of the electro-optic interconnect.

In an EOM, the modulation of the optical field is primarily subject to the electro-optic properties of the medium \cite{chr}. For EOMs based on the electro-optic effect, the refractive index $n$, i.e., the real part of the material susceptibility, is proportional to the electric field (Pockels effect) or the square of the electric field (Kerr effect) \cite{klc}. Although these well-studied effects have been exploited in numerous EOM devices, they require large components and/or a high driving voltage owing to the weakness of the electro-optic effect \cite{cpd} (e.g., $\Delta n\sim 3.3\times 10^{-3}$ for the Pockels effect in LiNbO$_3$ and $\Delta n\sim 10^{-5}$ for the Kerr effect in silicon with an applied field of $20 V/\mu m$ \cite{klc}). These requirements lead to higher energy dissipation than appropriate for off- or on-chip interconnection \cite{mda}.

Another modulation mechanism is the electro-absorption effects, in which the absorption spectra are modified by the applied electric field. Electro-absorption effects include the Franz-Keldysh effect \cite{mdab1} in bulk semiconductor materials and the quantum-confined Stark effect \cite{mdab2} in quantum well structures \cite{cpd,ljb}; these effects originate from the distortion of the energy bands by an applied electric field, which causes the modification of the absorption coefficient \cite{klc}. EOMs based on the electro-absorption effects, which typically involve III-V direct-band-gap semiconductors, are compatible with semiconductor technology \cite{cpd}, which lends itself to the development of compact devices with low power consumption \cite{lae,vjs}. For on-chip integration and miniaturisation, silicon-based EOMs are more desirable \cite{rgt}; however, their development is hampered by the weak electro-optic absorption effect of the indirect band gap of silicon \cite{cmp}. Thus a key area of EOM-related research is the investigation of novel materials with good electro-optic properties, such as a graphene-based broadband optical modulator with the several distinctive advantages including the strong light-graphene interaction, high-speed operation as well as
complementary metal-oxide-semiconductor compatible capability \cite{mlx}; the high performance GeSi electro-absorption modulator with the ultralow energy consumption offering unique advantages for use in high-performance electronic-photonic integration with complementary metal oxide semiconductor circuits \cite{cpd,lje}. Here, we report a new modulation mechanism for EOMs based on quantum interference, in which the refractive index and the absorption of the material are engineered coherently with unprecedented efficiency \cite{fmi}.

Quantum interference between two excitation pathways of the internal quantum states of a coherent three-level medium has given us the unique capability to engineer the linear and nonlinear susceptibility of the medium. In the case of electromagnetically induced transparency (EIT) \cite{kjb}, when the two metastable states couple to a common excited state by the control and probe fields, the absorption of the probe field in the medium is largely modified through the destructive quantum interference of the amplitudes of the two optical transitions. The width of the transparency window and the steepness of the refractive index curves are modulated over a wide range by dynamically changing the strength of the control field \cite{fmi}. This ability to modify the susceptibility has revealed many new phenomena in quantum optics, such as ultra-slow group velocities \cite{lvh} and the storage of light and quantum memory \cite{mfm}. Furthermore, this feature has been shown experimentally to be able to decrease the group velocity of light by many orders of magnitude  \cite{bdk,tav}. With the slowing of light, nonlinear optics can extend to the few-photon level \cite{bme,vvs}.

An optical switch based on quantum interference was proposed by Harris and Yamamoto in a four-level EIT atomic system \cite{seh}. Subsequently, this optical switching mechanism was experimentally demonstrated in a four-level $^{87}Rb$ atomic system \cite{mye}. In 2011, researchers have observed that the transmission of the probe field can be perfectly switched using cavity EIT and by applying a weak switching field \cite{maa}. These results demonstrate the ability of quantum interference to modulate an optical field.

Photons are the fastest and most robust information carriers. As such, the interconnects used in data communication systems and in long links will gradually change from electrical-based to light-based. However, photons must be slowed down to match the speed of the slow electrons in the electro-optic interconnect. Based on cavity-induced transparency, we previously proposed an approach to realise the electrical manipulation of the dark-state polariton and the group velocity of light \cite{lgqin}. Here, we propose a hybrid EOM system composed of a high-quality tunable cavity and an electro-mechanical system. By the interaction between a three-level medium confined inside the cavity and two optical fields, the group velocity of the modulated optical field in the medium is electrically slowed to match the slow electric signal. Through the use of electrically-driven quantum interference, we establish a one-to-one correspondence between the waveforms of the electrical signal and the optical field (that is, to create an electro-optic waveform interconnect) and use it to produce an arbitrary-waveform optical field. To the best of our knowledge, this work represents the first proposed electro-optic waveform interconnect based on such an approach.

Specifically, our scheme proposes implementing a coherent three-level medium that is confined in a compact device composed of a tunable cavity with a high quality factor and an electro-mechanical system, as shown in Fig. 1. The cavity consists of a moveable mirror, which is charged to operate as a charged mechanical oscillator (CMO), and a fixed mirror. The CMO has positive charge Q and capacitively couples to a fixed conductive plate that is supplied with negative charge -Q by a voltage waveform with time regime $T$ from a waveform generator. Thus the CMO and the fixed conductive plate form a mechanically variable capacitor. 

In this model, the control field injected into the cavity is a constant optical field with an amplitude $\varepsilon_c=\sqrt{2P_{c}\kappa/\hbar\omega_0}$, where the parameters represent respectively the frequency $\omega_0$, the power $P_c$, the Planck constant $\hbar$ and the cavity decay rate $\kappa$. In this structure, the displacement $q$ of the CMO, with a mass $m$ and frequency $\omega_m$, is driven by the radiation pressure force from the optical field and the Coulomb force from the capacitor. The Coulomb force and the radiation pressure forces on the CMO have the same direction and the Coulomb interactions can be approximated as $V_c\approx-U^2 \eta (q+r)$ for $q\ll r$, where $\eta=\frac{\varepsilon_0 S}{2 r^2}$, $\varepsilon_0$ is the vacuum permittivity, $U$ is the voltage, $S$ is the area and $r$ is the distance between two capacitive plates \cite{lgqin}. Because of the motion of the CMO, the resonance frequency in the cavity is modified as $\omega_c(q)=\omega_{c}(0)+q\partial\omega_c(q)/\partial q + O(q)$, where the high order terms $O(q)$ are neglected \cite{amt}. This process can lead to a detuning between the cavity field ($a$ ($a^{\dag}$) is the the annihilation (creation) operator of the photons in the cavity) and the control field $\Delta_{cmo}=\omega_c(q)-\omega_0$, which induces the leakage of photons in the cavity. The dynamical equations of the CMO and the cavity field are given by

\begin{figure}[h]
\includegraphics[angle=0,width=15cm]{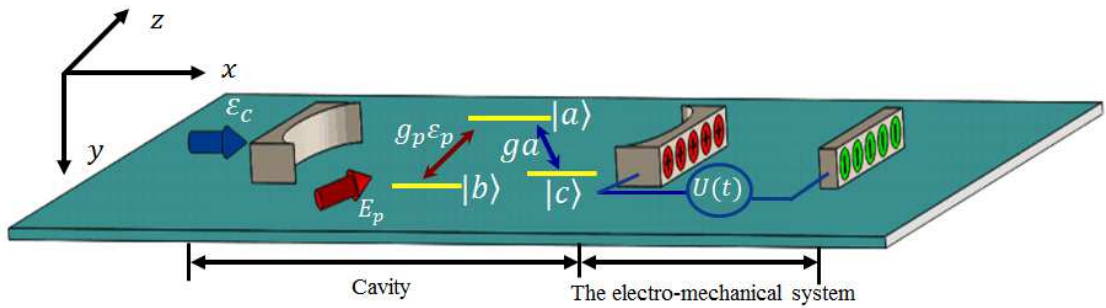}
\caption{(Color online)The proposed opto- and electro-mechanical hybrid system is composed of a tuneable cavity with a charged mirror operating as a CMO and a mechanically variable capacitor. A $\Lambda$-type three-level medium confined inside the cavity interacts with two optical fields: a constant optical field $\varepsilon_c$, which is resonantly injected into the cavity along the $x$ axis to form the cavity field, and the probe field $E_p$, which is externally injects into the cavity along the $z$ axis at frequency $\omega_p$.}
\end{figure}

\begin{eqnarray}
\label{eqgn1}
&&\ddot{q}+\gamma_m \dot{q}+ \omega^2_m q=(\hbar G_{0}n(t)+ U^2\eta)/m,\nonumber\\
&&\dot{a}=(-\kappa-i\Delta_{cmo})a+\varepsilon_c,
\end{eqnarray}
where $\gamma_m$ is the damping rate of the CMO, $n_c=a^{\dag}a$ is the photon number in the cavity and $G_0=-\partial\omega_c(q)/\partial q \approx\omega_c/L$ is the opto- and electro-mechanical coupling strength for a cavity of length $L$ \cite{amt}. Compared with the Coulomb force, the radiation pressure force can be neglected in our model and the motion of the CMO only depends on the Coulomb force. When the charges are slowly injected into the capacitor with a duration that satisfies $T\gg1/\gamma_m$ and $1/\kappa_{eff}$, the evolution of the displacement of the CMO and the cavity field can be considered adiabatically and the solutions of equation (1) can be derived as $q=\frac{U^2\eta}{m\omega_{m}^2}$ and $a=\frac{\varepsilon_{c}}{\kappa+i\Delta_{cmo}}$. When $\kappa\ll\Delta_{cmo}$, the photon number $n_c$ in the cavity is only controlled by the voltage, hence, the cavity field can be electrically and adiabatically switched. Therefore, we can define the effective cavity decay rate as $\kappa_{eff}=\sqrt{\kappa^2+\Delta_{cmo}^2}$ and quality factor of the cavity as $Q_{eff}=\omega_c/\kappa_{eff}$, which is dynamically modulated by changing the voltage on the capacitor.

To realise the modulation of a probe field by the voltage, i.e., EOM, an interactive medium with a three-level configuration must be confined at an antinode of the standing wave inside the cavity, in which the probe field couples with the atomic dipole transition between levels $|b\rangle$ and $|a\rangle$ with a coupling strength $g_p$ and detuning $\Delta_p$; and the quantised cavity field couples the atomic dipole transition between levels $|c\rangle$ and $|a\rangle$ with a coupling strength $g$ and detuning $\delta$. We assume that the probe field propagates in a direction perpendicular to that of the cavity field. The amplitude of the cavity field at this antinode depends only on time. Here we choose an ensemble of $N$ identical $\Lambda$-type cold atoms with a volume $V$ that interacts with the probe field and cavity field, as shown in Fig. 1, however, any types EIT-media can be used, such as $V$-type and the ladder type. The interaction Hamiltonian of the system is given by $H=-\hbar G_{0}a^{\dagger}aq+i\hbar\varepsilon_{c}(a^{\dagger}e^{-i\omega_{0}t}-ae^{i\omega_{0}t}) -\hbar N( g_{p}\varepsilon_{p}\sigma_{ab}+ ga\sigma_{ac} + H.c.)-U^2\eta(q+r)$, where $\sigma_{\alpha\beta}=\frac{1}{N}\sum_{i=1}^{N}|\alpha_i\rangle\langle\beta_i|$ are the slowly varying collective operators of the atomic ensemble.

The dynamic equations of the system can be simplified with the slowly varying amplitude and weak-field approximations \cite{mfm}. We then introduce the dimensionless slowly-varying field amplitude $E_p=\varepsilon_p\sqrt{\hbar\omega_p/2\epsilon_0 V}$ and assume that almost all atoms are in the ground state $|b\rangle$ and that the number of probe photons $n_p$ is considerably smaller than the number of atoms, i.e., $\epsilon=\sqrt{n_p/N}\ll 1$ (the weak-field approximation). Considering these approximations and keeping the zeroth-order terms of $\epsilon$, we determine that $\sigma_{bb}\approx1$, $\sigma_{aa}\approx\sigma_{cc}\approx\sigma_{ac}\approx0$ \cite{gnm} and that the remaining atomic evolutions $\sigma_{ba}$ and $\sigma_{bc}$ can be derived from
\begin{eqnarray}
\label{eqgn916}
\dot{R}=-MR+A,
\end{eqnarray}
where
\begin{eqnarray}
R=\left(
    \begin{array}{c}
      \sigma_{ba} \\
      \sigma_{bc} \\
    \end{array}
  \right),
  M=\left(
      \begin{array}{cc}
        \gamma+i\Delta_{p} & -iga \\
        -iga^{\dag} & \gamma_s+i(\Delta_{p}-\delta) \\
      \end{array}
    \right),
    A=\left(
        \begin{array}{c}
          i g_p \varepsilon_{p} \\
          0 \\
        \end{array}
      \right).
\end{eqnarray}

Here, $\gamma$ ($\gamma_s$) is the decay rate of the higher energy level $|a\rangle$ (the metastable state $|c\rangle$) to the ground state $|b\rangle$ with $\gamma_s\ll\gamma$. For a sufficiently long time, the solution of equation (\ref{eqgn916}) can be generally provided by
\begin{eqnarray}
\label{eqgn11101}
R=e^{-Mt}R_0+(1-e^{-Mt})M^{-1}A\approx M^{-1}A.
\end{eqnarray}
This yields
\begin{eqnarray}
\label{eqgn2}
\sigma^s_{ba}=\frac{i g_p\varepsilon_p[\gamma_s+i(\Delta_p-\delta)]}{(\gamma+i\Delta_p)[\gamma_s+i(\Delta_p-\delta)]+g^2(n_c+1)}.
\end{eqnarray}

Thus the first-order susceptibility of the medium can be derived as $\chi=\frac{\mu_{ba} N}{\epsilon_0 E_p V}\sigma_{ba}^s$, where $\mu_{ba}$ is the dipole moment of the probe transition. The imaginary part of this susceptibility, $Im(\chi)$, determines the dissipation of the probe field, i.e., absorption, whereas the real part, $Re(\chi)$, determines the refractive index \cite{fmi}. Equation (\ref{eqgn2}) indicates that the width of the transparency window ($\Delta\omega_{EIT}$) and the dispersion of the probe field depend on the intensity of the control field ($g^2(n_c+1)$) \cite{jao}, which is regulated by the voltage. In Fig. 2, we observe that the width of the transparency window decreases and even disappears when the voltage $U^2$ increases. In this process, the probe field is gradually absorbed by the medium and can even be completely absorbed with a large voltage. Meanwhile, the slope of the steep dispersion is also significantly modified by the voltage (Fig. 2 inset). Because we use a high-Q cavity and cavity induced transparency, the strong control field required in conventional EIT \cite{fmi} has been reduced to an assistant optical field with a low and constant intensity, that is, for the observable features of EIT, provided that the cavity field satisfies $g^2(n_c+1)\gg\gamma\gamma_s$. Based on the mechanism of cavity-induced transparency \cite{gnm}, quantum interference can be then manipulated by the electric field in this integrated opto- and electro-mechanical system. In the following we will show that the absorptive waveform of the modulated probe field has a one-to-one correspondence with the waveform of the voltage.

For the non-resonant case $\Delta_p\neq0$, the width of the transparency window determines the bandwidth of the modulation, which will be discussed later. Now to simplify the expression of $\chi$, we assume that the system operates are in the resonant regime, i.e., $\Delta_p=\delta=0$. The imaginary part of the susceptibility is then given by
\begin{eqnarray}
\label{eqgn3}
Im(\chi)^s=\chi_0\frac{\gamma_s}{\gamma\gamma_s+g^2[\frac{\varepsilon^2_c}{\kappa^2+(\frac{2G_0\eta}{m w_{m}^2}U^2)^2}+1]},
\end{eqnarray}
where $\chi_0=\frac{\mu^{2}_{ba} N}{\hbar\epsilon_0 V}$. Thus we obtain an analytic expression of the relationship between the absorption of the medium $Im(\chi)$ and the voltage $U^2$. When $U^2=0$, the photon number in the cavity field is maximum, described by $n_{cmax}=\frac{|\varepsilon_c|^2}{\kappa^2}$. The strong cavity field induces a probe field transparent in the coherent medium. From equation (\ref{eqgn3}), the minimum absorption in the transparency window is given by $Im(\chi)_{min}=\frac{\chi_{0}\gamma_{s}}{g^2(n_{cmax}+1)+\gamma\gamma_{s}}$. In real atomic systems, because the coherence decay rate $\gamma_s$ with the forbidden $|b\rangle$ to $|c\rangle$ transition is nonzero, the transparency is never perfect. When $\gamma\gamma_s\ll g^2(n_{cmax}+1)$, we can consider that the medium is transparent. When $n_c$ approaches zero for a large voltage, the maximum absorption is $Im(\chi)_{max}=\frac{\chi_{0}\gamma_{s}}{g^2+\gamma\gamma_{s}}$. The inverse transformation of the corresponding square of the voltage as a function of $Im(\chi)$ can also be easily derived from equation (\ref{eqgn3}) as
\begin{eqnarray}
\label{eqgn5}
U^2=\frac{m\omega^{2}_{m}}{2G_{0}\eta}\sqrt{\frac{g^2\varepsilon^{2}_{c}}{\frac{\chi_0\gamma_s}{Im(\chi)^s}-\gamma\gamma_s-g^2}-\kappa^2}.
\end{eqnarray}
where $\frac{\chi_{0}\gamma_{s}}{g^2(n_{cmax}+1)+\gamma\gamma_{s}}\leq Im(\chi)^s<\frac{\chi_{0}\gamma_s}{\gamma\gamma_s+g^2}$. From equations (\ref{eqgn3}) and (\ref{eqgn5}), we can find a one-to-one correspondence between the voltage and the absorption of the probe field, that is, the change of the voltage can lead to the modulation of the absorption of the probe field. Therefore the EOM based on the electro-optic absorption is achieved. Using this interconnect, we can generate an arbitrary waveform of the optical field by adjusting the corresponding voltage waveform.

\begin{figure}
\includegraphics[angle=0,width=15cm]{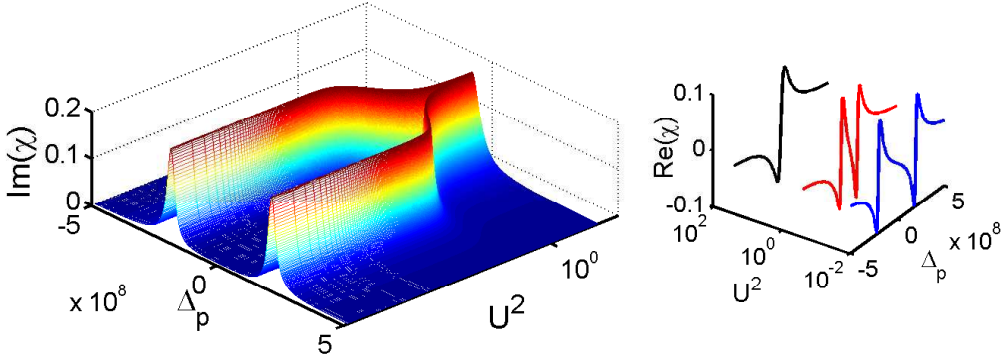}
\caption{Imaginary part of the susceptibility of the probe field in the medium as a function of the square of the voltage $U^2$ and the detuning $\Delta_p$. The inset shows the real part of the susceptibility. Here the units of the voltage and the detuning axis are the square of volt ($V^2$) and the hertz ($Hz$), respectively. The parameters are used from the experiments in Ref.\cite{sgk} as $\varepsilon_c=4\times10^{10}Hz;$ $\gamma=2\pi\times5.75$MHz; $\gamma_s=0.0001\gamma; g=0.001\gamma; \kappa=0.2\gamma; \omega_m=\gamma; m=145$ng; $G_0=2\pi \times 1.5\times10^{16}Hz/m; S=0.6 mm^2; r=0.21\mu m$; $\delta=0$ and atomic density $\sim 10^{19} m^{-3}$. The modulative material is the $^{87}$Rb with the $\Lambda$-type three levels configuration.}
\end{figure}

If the change of the cavity field caused by the voltage satisfies the adiabatic conditions described in references \cite{mfm,gnm,fml}, the electrically controlled dark-state polariton and group velocity can be obtained as \cite{lgqin}
\begin{eqnarray}
\label{eqgn11104}
&&\Psi=\cos\theta\varepsilon_{p} - \sin\theta \sqrt{N}\sigma_{bc},\nonumber\\
&&\nu_g= c\;\cos^2\theta,
\end{eqnarray}
where the mixed angle $\theta$ is defined as $\tan^2\theta=\frac{g_{p}^{2} N}{g^2(\hat{n}+1)}$ and depends on the voltage on the capacitor. Equation (\ref{eqgn11104}) shows that the properties of the dark-state polariton and the group velocity are determined by the mixing angle $\theta$ between the probe field and the atomic spin states. By modulating the voltage, the mixing angle can be changed in the range $0-\pi/2$, which can lead to store and retrieve the probe field in the medium, and the slowing of the group velocity. Therefore the absorption rate and the group velocity of the probe field are coherently and simultaneously modulated.

To demonstrate the ability of the proposed scheme to modulate the optical field, we give the simulation results. For the modulative waveforms, we choose that the standard and general sine, sawtooth and square waveforms are the target waveforms, shown by the solid lines in Figs. 3(a1), (b1) and (c1). In order to obtain the regular target waveforms, the corresponding voltage waveforms of Figs. 3(a2), (b2) and (c2) with the amplitude $U_m$ were applied to the capacitor by the waveform generator. Then, using numerical simulations, we obtained the corresponding modulated waveforms, indicated by the dashed lines in Figs. 3(a1), (b1) and (c1). The differences between the target and the numerical results mainly originate from the transient properties of the EIT, which can lead to negative absorption, that is, the transient gain \cite{lyq,srd}. The gains
appear in the transparency window of the probe, therefore the modulation amplitude can be enhanced. The numerical results show that the modulated absorptive sine, sawtooth and square waveforms all can follow the voltage waveform, and that an arbitrary-waveform modulator can be produced based on our approach.

\begin{figure}
\includegraphics[angle=0,width=20cm]{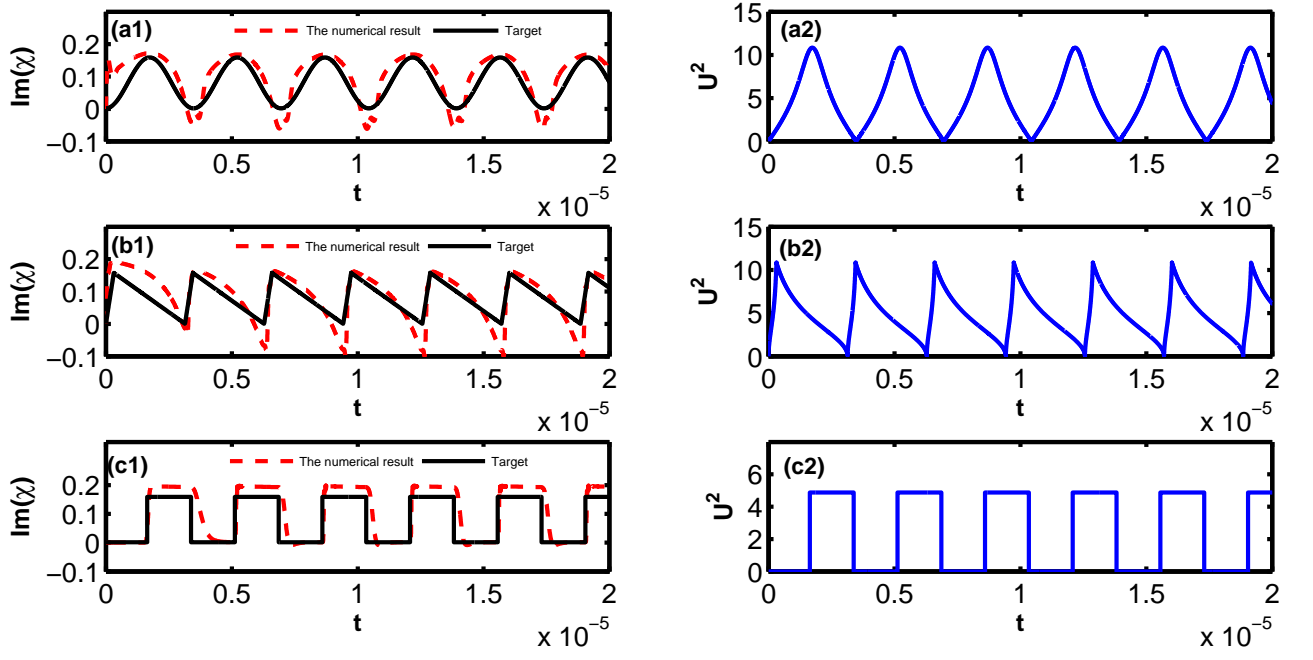}
\caption{Numerical results of the EOM. (a) The modulation of the sine wave: (a1) the target absorptive waveform and numerical results, (a2) the square of voltage waveform $U^2$ applied to the capacitor. (b) the modulation of the sawtooth wave: (b1) The target absorptive waveform and numerical results, (b2) The square of voltage waveform $U^2$ applied to the capacitor. (c) the modulation of the square wave: (c1) The target absorptive waveform and numerical results, (b2) The square of voltage waveform $U^2$ applied to the capacitor. The simulation parameters are $\varepsilon_c=0.5\times10^{10}Hz$, $\gamma_m=3\gamma$ and $\kappa=0.4\gamma$, The other parameters are same in Fig. 2. }
\end{figure}

The extinction ratio ( also known as the modulation depth) is used to describe the modulation efficiency of the transmitted light and is defined as the ratio of the maximum transmission $I_{max}$ to the minimum transmission $I_{min}$, in other words $10log(I_{max}/I_{min})$ \cite{rgt}. To measure the efficiency of the electrically controlled absorption, we can define the extinction ratio as
\begin{eqnarray}
\label{eqgn7}
R_{dB}=10log\frac{Im(\chi)_{max}}{Im(\chi)_{min}},
\end{eqnarray}
where the maximum absorption $Im(\chi)_{max}$ is obtained without the cavity field when the voltage is maximum; the minimum absorption $Im(\chi)_{min}$ corresponds to the strongest cavity field, which is obtained when the voltage is zero. From the analytical equation (\ref{eqgn3}), the extinction ratio is calculated as $R_{dB}=10log \frac{\gamma\gamma_s+g^2(n_{cmax}+1)}{\gamma\gamma_s+g^2(n_{cmin}+1)}$, where $n_{cmin}=\frac{\varepsilon^2_c}{\kappa^2+(\frac{2G_0\eta}{m w_{m}^2}U_{m}^2)^2}$. Therefore we can enhance the extinction ratio by increasing the difference between $n_{cmax}$ and $n_{cmin}$.

The modulation speed of the EOM depends on the following factors. First, the motion of the CMO must follow the change of the voltage, that is, it is operates as a driven damped mechanical oscillator. The frequency of the oscillation follows the frequency of the strong driving field and its amplitude and phase are determined by the mass, frequency and damping of CMO and by the frequency and the amplitude of the strong driving field \cite{wwn}. Second, the cavity field is switched dynamically by the CMO based on the cavity detuning. Therefore, the quality factor of the cavity must match the cavity detuning, whose magnitude can be modulated by the area and distance of the two capacitive plates. For high frequency modulation, the nonadiabatic changes of the cavity field are caused by the difference between the quick mechanical detuning and the leakage of the cavity itself. Third, for a high-frequency driving voltage, the interaction between the optical fields and the medium can occur under nonadiabatic conditions and the nonadiabatic transfer of the dark-state polariton between the probe field and the atomic spin state can be obtained, as discussions in \cite{mab}. Therefore, a high frequency of modulation is possible when the above conditions are met. In addition, the modulation bandwidth, the range of the probe frequency in the transparency window ($\Delta_p$), depends on the strength of the cavity field, which can be increased by improving the quality factor and strengthening the optical field $\varepsilon_c$.

In the proposed scheme, it is a key that the absorption rate in the EIT can be modulated by controlling electrically the control field based on quantum interference, therefore any coherent medium with an energy level configuration that enables EIT can serve as the modulation medium. The energy level configuration can be extended from $\Lambda$-type to $V$ - and cascade-type three-level systems, even to general $N$- \cite{ykn} or $M$- type \cite{dcl} configurations. Moverover, the medium could include a wide range of systems, for example, a semiconductor quantum dot \cite{bdj} or quantum well \cite{wzy}, etc.

Compared to conventional EOM technology, which is based on the electro-optic properties of the constituent materials, our proposed quantum-interference-based EOM to form electric-optic interconnect offers a number of advantages. First, the steep EIT-induced dispersion can lead to a modification of the refractive index over a wide range and even change the refractive index from positive to negative \cite{omo,tqm}. Second, the absorption rate can be greatly modulated by adjusting the transparency window, in which the complete absorption and transparency can be switched. Third, in the process of the coherent interaction, the refractive index and absorption rate are modulated simultaneously. Fourth, in principle, the proposed scheme can overcome the dependence of conventional EOMs on the electro-optic properties of materials.

In conclusion, we propose a new mechanism to realise an electro-optic waveform interconnect by EOM based on quantum interference. The proposed new mechanism maybe open a new field to study EOM and the waveform transfer from the electrical signal to the optical one. Owing to quantum interference mechanism as one of the core in quantum information, the electro-optic modulation, especially, the electro-optic waveform transfer, may be used in the photonic modulation at the quantum level. This new mechanism exhibits unprecedented capacity for EOM in the waveform electro-optic interconnects and maybe have potential applications in the future communication and signal processing systems.

\noindent\textbf{Acknowledgements}\\
We acknowledge Dr. H. Shen, J. Lu and Simon-Pierre Gorza for language editing and programming supper. This work is supported by the Strategic Priority Research Program (No. XDB01010200), the Hundred Talents Program of the Chinese Academy of Sciences (No. Y321311401), National Natural Sciences Foundation of China (Nos. 11347147, 11274112, 11547035 and 91321101) and Shanghai Advanced Research Institute (Grant 141004).

\noindent\textbf{Author contributions}\\
L. G. Qin and Z. Y. Wang contributed to the conception and design of this work. S. Q. Gong participated in the discussion of the results. All authors contributed in the article preparation. Z. Y. Wang supervised this project.

\noindent\textbf{Additional Information} \\
Competing financial interests: The authors declare no competing financial interests.

\end{document}